\documentstyle{mn}

\newif\ifAMStwofonts

\ifoldfss
  \ifCUPmtlplainloaded \else
    \NewTextAlphabet{textbfit} {cmbxti10} {}
    \NewTextAlphabet{textbfss} {cmssbx10} {}
    \NewMathAlphabet{mathbfit} {cmbxti10} {} 
    \NewMathAlphabet{mathbfss} {cmssbx10} {} 
  \fi
  \ifAMStwofonts
    \ifCUPmtlplainloaded \else
      \NewSymbolFont{upmath} {eurm10}
      \NewSymbolFont{AMSa} {msam10}
      \NewMathSymbol{\upi}     {0}{upmath}{19}
      \NewMathSymbol{\umu}     {0}{upmath}{16}
      \NewMathSymbol{\upartial}{0}{upmath}{40}
      \NewMathSymbol{\leqslant}{3}{AMSa}{36}
      \NewMathSymbol{\geqslant}{3}{AMSa}{3E}

      \let\leq=\leqslant 
      \let\geq=\geqslant 
    \fi
  \fi
\fi 

\ifnfssone
  \newmathalphabet{\mathit}
  \addtoversion{normal}{\mathit}{cmr}{m}{it}
  \addtoversion{bold}{\mathit}{cmr}{bx}{it}
  \newmathalphabet{\mathbfit} 
  \addtoversion{normal}{\mathbfit}{cmr}{bx}{it}
  \addtoversion{bold}{\mathbfit}{cmr}{bx}{it}
  \newmathalphabet{\mathbfss} 
  \addtoversion{normal}{\mathbfss}{cmss}{bx}{n}
  \addtoversion{bold}{\mathbfss}{cmss}{bx}{n}
  \ifAMStwofonts
    \ifCUPmtlplainloaded \else
      \UseAMStwoboldmath
      \makeatletter
      \new@mathgroup\upmath@group
      \define@mathgroup\mv@normal\upmath@group{eur}{m}{n}
      \define@mathgroup\mv@bold\upmath@group{eur}{b}{n}
      \edef\UPM{\hexnumber\upmath@group}
      \new@mathgroup\amsa@group
      \define@mathgroup\mv@normal\amsa@group{msa}{m}{n}
      \define@mathgroup\mv@bold\amsa@group{msa}{m}{n}
      \edef\AMSa{\hexnumber\amsa@group}
      \makeatother
      \mathchardef\upi="0\UPM19
      \mathchardef\umu="0\UPM16
      \mathchardef\upartial="0\UPM40
      \mathchardef\leqslant="3\AMSa36
      \mathchardef\geqslant="3\AMSa3E

      \let\leq=\leqslant 
      \let\geq=\geqslant 
    \fi
  \fi
\fi 

\ifnfsstwo
  \DeclareMathAlphabet{\mathbfit}{OT1}{cmr}{bx}{it}
  \SetMathAlphabet\mathbfit{bold}{OT1}{cmr}{bx}{it}
  \DeclareMathAlphabet{\mathbfss}{OT1}{cmss}{bx}{n}
  \SetMathAlphabet\mathbfss{bold}{OT1}{cmss}{bx}{n}
  \ifAMStwofonts
    \ifCUPmtlplainloaded \else
      \DeclareSymbolFont{UPM}{U}{eur}{m}{n}
      \SetSymbolFont{UPM}{bold}{U}{eur}{b}{n}
      \DeclareSymbolFont{AMSa}{U}{msa}{m}{n}
      \DeclareMathSymbol{\upi}{0}{UPM}{"19}
      \DeclareMathSymbol{\umu}{0}{UPM}{"16}
      \DeclareMathSymbol{\upartial}{0}{UPM}{"40}
      \DeclareMathSymbol{\leqslant}{3}{AMSa}{"36}
      \DeclareMathSymbol{\geqslant}{3}{AMSa}{"3E}

      \let\leq=\leqslant 
      \let\geq=\geqslant 
    \fi
  \fi
\fi 

\ifCUPmtlplainloaded \else
  \ifAMStwofonts \else 
    \def\upi{\pi}
    \def\umu{\mu}
    \def\upartial{\partial}
  \fi
\fi

\title{Copper and Barium Abundances in the Ursa Major Moving Group}
\author[S. Castro et al.]
       {S. ~Castro,$^1$
       G. F. ~Porto de Mello,$^1$ L. ~da Silva$^2$\\
       $^1$Observat\'orio do Valongo/UFRJ, Ladeira do Pedro Ant\^onio, 43,
       20080-090, Rio de Janeiro, RJ, Brazil\\
       $^2$Observat\'orio Nacional,Departamento de Astronomia, Rua
       Gen. Jos\'e Cristino, 77, 20921-400, Rio de Janeiro, Brazil}
\date{Accepted-----------
      Received-----------}

\pagerange{\pageref{firstpage}--\pageref{lastpage}}
\pubyear{1998}

\begin{document}

\maketitle

\label{firstpage}

\begin{abstract}
We present Cu and Ba abundances for 7 G-K dwarf
stars, members of the solar-metallicity, 0.3 Gyr old Ursa Major Moving
Group. All analyzed member stars show [Ba/Fe] excesses of +0.3-plus,
associated with [Cu/Fe] deficiencies of up to $-$0.23 dex. The present results
suggest that there is an anti-correlation between the abundances of
Cu and the heavy elements produced by the main component of
the neutron capture s-process. Other possible
anomalies are Na and C deficiencies with
respect to normal solar-metallicity stars. The new data do
not confirm the recent claim that the group member HR6094 is a Ba 
dwarf star.

\end{abstract}


\input epsf.sty

\section{Introduction}
Associations and clusters are very good samples of stars which
provide information on the chemical composition of homogeneous
stars whose physical parameters can be determined with reduced uncertainties
as compared to field stars. There is not a nearby, well-populated
open cluster with an age 0.2 -- 0.3 Gyr old that can be analyzed in
detail such as the Pleiades and Hyades, although the Ursa Major Group
(UMaG hereafter) is a loose and sparse group of stars that kinematically reveal 
their common origin. The UMaG is a 0.3 Gyr old
kinematical group of solar metallicity located in the solar neighborhood.
Soderblom \& Mayor (1993) identified probable member stars of the 
group as being those stars where
both the kinematical parameters and the chromospheric
activity indicators point decidedly towards membership. 
As pointed out by Soderblom \& Mayor, there are good reasons
for the study of UMaG in detail. The UMaG has a sparse but
tangible nucleus in the velocity space; statistically, 
it appears to be real; the
kinematics of the UMaG stars are well determined due to
the group's proximity; there are reliable spectroscopic
indicators to estimate the ages of young solar-type
stars; its space velocity is distinct from other young
field stars so that there should be few interlopers to
contaminate the sample. Therefore
the UMaG is important in the study of stellar kinematic groups as well as
feasible in providing a sample of young stars not found in
other nearby clusters.

Porto de Mello \& da Silva (1997) claimed the solar-type, solar
metallicity star HR6094, member of the UMaG,
to be the first solar-metallicity,
young Ba dwarf star to be identified. They found marked
overabundances for the s-process elements which could be explained
by the process of mass transfer in a binary system, in which the
secondary component accreted matter from the primary one (now the
white dwarf companion) when it was an asymptotic giant branch star
self-enriched in s-process elements. A 30 Myr old, common proper
motion white dwarf star located 0.026 pc away was tentatively
identified as the former primary: if confirmed, this could be
the first Ba system in which the remnant of the late AGB star
responsible for the heavy elements enrichment may have been
directly spotted.
If so, this would be
the first identification of the Ba phenomenon in a near-zero-age
star of solar metallicity. We have observed seven G-K dwarf stars (including 
HR6094), selected by Soderblom \& Mayor (1993) as probable members of the UMaG,
in order to establish the abundance pattern of the UMaG stars and therefore 
the anomalous status of HR6094 within the UMaG, thereby confirming
its condition of a Ba dwarf star. Besides this, we aim to study
the possible existence of a connection between the abundances of Cu and
Ba.

The studies of Pereira \& Porto de Mello (1997) and Pereira, Smith \& Cunha 
(1998) verified that there is a deficiency of Cu abundances relative
to Fe connected to an overabundance of [Ba/Fe] in symbiotic stars.
Another similar result is found by Vanture (1992) for the CH star
HD26, which shows [Cu/Fe]=$-$1.5 and [Y/Fe]=$+$1.0 dex.
The scientific importance in studying the abundance of Cu is because
it has been noticed that Cu is manufactured by several mechanisms still
poorly known in the literature. The study of [Cu/Fe] in stars of different
metallicities gives an important indication of the nucleosynthetic
processes derived from the progenitor stars that contribute to the
formation of this element. The astrophysical site for the synthesis of
Cu is not yet well established. The observational study of Sneden, Gratton \&
Crocker
(1991) concludes that the behavior of Cu in a sample
of stars in the metallicity interval $-$2.9 $\leq$ [Fe/H] $\leq$ $-$0.1
may be understood if it is mainly produced through the s-process
weak component in massive stars with a small contribution by
explosive burning Type II Supernovae. On the other hand, the theoretical
study of Matteucci {\it et al.} (1993) fits their best Galactic chemical
evolution model to the available observations and shows that the
evolutionary histories of Cu is compatible with a major
contribution ascribed to Type Ia Supernovae, by means of a primary
process. The nucleosynthesis of Ba is rather well established to be  
produced at the 89 per cent level by the s-process main component in AGB stars of low masses.
Based on these two scenarios, we have performed a detailed abundance
analysis of the elements Fe, Cu and Ba in a sample of seven UMaG member
stars and have compared their abundances with those of normal disc
stars as described in the next sections.

\section[]{Observations and Analysis}

High S/N, moderately high-resolution spectra of the UMaG stars and a
sample of normal solar-type disc stars were obtained at the coud\'e
spectrograph of 1.6-m telescope of the Observat\'orio do Pico dos
Dias, Brazil. We discuss seperately the observations for the two
samples.

The selected UMaG stars were observed in two 140 \AA\ wide spectral
regions centered in $\lambda$5810 and $\lambda$6145, in September
1997. We targeted at the $\lambda$5853, $\lambda$6141 Ba II and the
$\lambda$5782 Cu I spectral features. Na I, Fe I and Fe II lines are
also present in the $\lambda$6145 region; a few useful Fe I lines were
measured in the $\lambda$5810 region. The spectral resolution was
0.30 \AA\ and all spectra were integrated to S/N $\geq$ 200. Three of
the seven selected UMaG stars were observed in both regions: the
remaining were observed only in $\lambda$5810. Clear sky spectra were
obtained in both regions as a solar template.

A sample of 13 solar-type disc stars spanning -0.8 $<$ [Fe/H] $<$ 0.0
were observed between 1991 and 1997 in 140 \AA\ wide spectral regions centered at
$\lambda$5244 and $\lambda$5810, at a resolution of 0.3 \AA. Three
more 100 \AA\ wide spectral regions centered at $\lambda$6053,
$\lambda$6145 and $\lambda$6707 were observed at a resolution of 0.2
\AA. S/N ratios in excess of 200 were obtained for nearly all stars.
There are many unblended Fe I and Fe II lines in these spectral
ranges: we also targeted at the $\lambda$5218 Cu I line. Sky and Moon
spectra were observed at very high S/N for each region.

Effective temperatures for the UMaG stars were determined from the
(B-V), (b-y) and $\beta$ color indices through the calibrations of
Saxner \& Hammarb\"ack (1985). Most of the UMaG stars had no Fe II lines for
a spectroscopic determination of the surface gravity. This parameter
was then determined by plotting the stars in the theoretical
solar-metallicity HR diagram of Charbonnel {\it et al.} (1993) by means of
the derived T$_{\rm eff}$s, Hipparcos paralaxes (ESA, 1997) and the
bolometric corrections of Habets \& Heintze (1981). Surface gravities
were calculated by deriving the stellar masses from their positions
with respect to the theoretical tracks. Microturbulent velocities
$\xi$ were obtained by forcing the Fe I line abundances to be
independent of line strength: this was done only for HR2047, HR6094
and HR6748, observed in both the $\lambda$5810 and $\lambda$6145
regions. For all other stars $\xi$ was derived from the relation given
by Edvardsson {\it et al.} (1993) (hereafter, E93). Atmospheric parameters
for the solar-type disc star sample were determined exclusively from
the spectral data. Effective temperatures, surface gravities, Fe
abundances and microturbulent velocities were determined from the
simultaneous solution of the excitation \& ionization equilibria of Fe.
Exceptions to this are HR2883, HR3018, HR7875, HR8181, and HR8697 for
which the atmospheric parameters were taken from the E93 analysis. In all
cases, the Fe abundances derived from the Fe I lines available
in our spectra were fully consistent with the metallicities given
by E93.

The equivalent widths (W$_\lambda$s) of the spectra were converted
to elemental abundances using the modified MARCS atmospheric models
(E93) and a computer code developed by M. Spite (Paris-Meudon
Observatory). Tables 1 and 2 show the measured W$_\lambda$s for
all available lines in the UMaG stars and normal disc stars,
respectively. Both the stellar and solar W$_\lambda$s were
converted to the W$_\lambda$ system of Meylan, Furenlid \& Wiggs (1993), who
generated solar W$_\lambda$s by fitting Voigt profiles to the
observed line profiles of the Kurucz {\it et al.} (1984) Solar Atlas, by
means of the relation W$_\lambda$(OPD) = 1.05  W$_\lambda$
(Atlas). This relation has been established by a linear regression between the
common lines, and is consistent with the low scattered light levels
expected for this conventional coude spectrograph. The dispersion of
the regression is 3 m\AA\ and we take this as an estimate of the error
of the W$_{\lambda}$ measurements. Solar {\it gf} values were derived from the 
solar W$_\lambda$s
and a solar model atmosphere also computed in the modified MARCS code.
For the $\lambda$5218 and $\lambda$5782 Cu I lines, the hyperfine
structure was explicitly taken into account according to the data of
Steffen (1985) (Table 3).

All analyses were rigorously differential with respect to the Sun, the
standard star, for which we adopted T$_{\rm eff}$ = 5777 K, log g =
4.44 and $\xi$ = 1.2 km.s$^{\rm -1}$. In such a differential analysis
the internal errors are the ones to worry about: these are estimated
at 1$\sigma$, respectively, for the UMaG stars and solar-type disc
stars, as $\sigma$(T$_{\rm eff}$) = 70 K, $\sigma$(log g) = 0.10 dex,
$\sigma$ = ([Fe/H]) = 0.12 dex, $\sigma$($\xi$) = 0.3 km.s$^{\rm -1}$,
and $\sigma$(T$_{\rm eff}$) = 70 K, $\sigma$(log g) = 0.30 dex,
$\sigma$ = ([Fe/H]) = 0.07 dex, $\sigma$($\xi$) = 0.153 km.s$^{\rm
-1}$. For the disc star sample with the exception of the stars
for which the atmospheric parameters were taken from E93, we opted for
the spectroscopic surface
gravities to keep with atmospheric parameters derived exclusively from
the line data: an analysis in the theoretical HR diagram similarly as
done for the UMaG stars pointed to full consistency between the
spectroscopic surface gravities and the gravities derived from the
Hipparcos parallaxes. Atmospheric parameters and elemental abundances
are given, respectively, for the UMaG stars and solar-type disc stars,
in Tables 4 and 5. In Table 4 one may notice the good consistency,
within the uncertainties, of the [FeI/H] and [FeII\H] abundances.
The abundance data sets of the UMaG and the disc stars are thus fully
consistent with each other, allowing direct comparisons.

In Table 6 we list the effects upon
the abundance ratios [Cu/Fe] and [Ba/Fe] of changing the atmospheric
parameters and the W$_{\lambda}$ within the estimated errors. Errors
in T$_{\rm eff}$ and in the W$_{\lambda}$s are seen to dominate the
total uncertainty of the [Cu/Fe] ratio, which is 0.09 dex, whereas
errors in the microturbulent velocity mostly affect [Ba/Fe], for which
the compounded r.m.s. uncertainty is 0.11 dex

\begin{table*}
  \caption{Measured equivalent widths for the Ursa Major stars.}
  \begin{tabular}{@{}lcccccccccc@{}} 
&&&&\multicolumn{7}{c}{W$\rm {_\lambda(\AA)}$} \\[10pt]
Ion&$\rm {\lambda(\AA)}$& $\chi_{ex}$& log{\it gf}&
HR531B & HR1321 & HR1322 & HR2047 & HD41593 & HR6094 & HR6748 \\[10pt]
FeI&5778.463&2.59& -3.44&25 &21&30&---&43&24&--- \\
FeI&5811.916&4.14& -2.41&---&9 &15 &10 &17 &---&--- \\
FeI&5814.805&4.28& -1.82&---&22&28 &---&---&26 &22 \\
FeI&5852.222&4.55& -1.18&40 &36&45 &42 &61 &42 &34 \\
FeI&5855.086&4.61& -1.53&23 &---&27 &19&32 &23 &18 \\
FeI&5856.096&4.29& -1.54&33 &30&40 &31 &49 &38 &31 \\
FeI&5859.596&4.55& -0.42&79 &71&86 &76 &96 &80 &74 \\
FeI&6093.649&4.61& -1.35&---&---&---&30&---&33 &28 \\
FeI&6096.671&3.98& -1.80&---&---&---&34&---&40 &--- \\
FeI&6151.623&2.18& -3.27&---&---&---&40&---&51 &42 \\
FeI&6159.382&4.61& -1.89&---&---&---&---&---&17 &--- \\
FeI&6173.340&2.22& -2.79&---&---&---&66 &---&74 &64 \\
FeI&6187.995&3.94& -1.60&---&---&---&42 &---&50 &42 \\
FeI&6191.571&2.43& -1.56&---&---&---&---&---&138&--- \\
FeII&6084.105&3.20& -3.81&---&---&---&26&---&23 &19 \\
FeII&6149.249&3.89& -2.72&---&---&---&42&---&43 &38 \\
CuI&5782.136&1.64& ---&65 &55 &73 &57 &95 &67 &54 \\
BaII&5853.688&0.60& -0.76&80&78 &85 &82 &83 &89 &84 \\
BaII&6141.727&0.70& +0.30&---&---&---&152&---&161&151 \\
\end{tabular}
\end{table*}

\vspace{1cm}

\begin{table*}
  \caption{Measured equivalent widths for the normal disc stars.}
  \begin{tabular}{@{}lccccccccccccccccc@{}} 
&&&&\multicolumn{13}{c}{W$\rm {_\lambda(\AA)}$} \\[10pt]
Ion&$\rm {\lambda(\AA)}$& $\chi_{ex}$& log{\it gf}&Moon&HR&HR&HR&HR&HR&HR&HR&
HR&HR&HR&HR&HR&HR\\
&&&&&77 & 98 & 173 & 509 & 695 & 914 & 2883 & 3018 & 7875 & 8181 & 
8501 & 8697 & 9088\\[10pt]

FeI  & 5196.065 & 4.26 & -0.72 &  81 &  65 &  73 &  70 &  80 &  76 & 
75 &  41 & --- &  55 & --- & --- & --- & --- \\ 
FeI  & 5197.929 &
4.30 & -1.51 &  37 &  24 &  31 &  26 &  26 &  34 &  28 &  12 &  10 & 
18 & --- & --- & --- & --- \\ 
FeI  & 5223.188 & 3.63 & -2.14 &  37 &
 24 &  32 &  21 &  25 &  36 &  33 &   7 &  10 &  16 &   6 &  23 &  15&  14 \\ 
FeI  & 5225.525 & 0.11 & -4.37 &  88 &  68 &  82 &  81 & 
82 &  85 & 100 &  39 &  44 &  53 &  30 &  67 &  45 &  63 \\ 
FeI  &5242.491 & 3.63 & -1.04 &  93 &  79 &  88 &  78 &  85 &  93 &  85 & 
54 &  56 &  72 &  53 &  80 &  74 &  66 \\ 
FeI  & 5243.773 & 4.26 &
-0.95 &  68 &  53 &  62 &  51 &  58 &  70 &  54 &  27 &  29 &  45 & 
27 &  55 &  46 &  39 \\ 
FeI  & 5247.049 & 0.09 & -4.57 &  81 &  60 &
 77 &  78 &  81 &  80 &  99 &  33 &  40 &  56 &  28 &  65 & --- &  61\\ 
FeI  & 5250.216 & 0.12 & -4.54 &  81 &  59 &  74 &  75 &  80 & 
76 &  93 &  32 &  37 &  48 &  27 &  65 &  39 &  60 \\ 
FeI  & 5778.463
& 2.59 & -3.44 &  25 &  15 &  17 &  18 &  20 & --- &  24 & --- & ---
&  11 & --- &  16 &  10 & --- \\ 
FeI  & 5811.916 & 4.14 & -2.41 & 
11 & --- & --- & --- &   6 & --- & --- & --- & --- & --- &   6 & --- &
--- & --- \\ 
FeI  & 5814.805 & 4.28 & -1.82 &  25 &  15 &  22 & ---
&  16 &  23 &  24 & --- & --- &  11 & --- &  17 &  11 & --- \\ 
FeI  &5852.222 & 4.55 & -1.18 &  43 &  29 &  35 &  32 &  32 &  40 &  34 &
--- & --- &  19 &   9 &  29 &  22 &  22 \\ 
FeI  & 5855.086 & 4.61 &
-1.53 &  24 &  14 &  18 &  14 &  18 &  21 &  14 & --- & --- &  11 &
--- &  18 &  11 &   8 \\ 
FeI  & 5856.096 & 4.29 & -1.54 &  37 &  24
&  30 &  25 &  27 &  33 &  32 & --- & --- &  16 & --- &  26 &  16 &--- \\ 
FeI  & 5859.596 & 4.55 & -0.42 &  86 &  62 &  71 &  56 &  64
&  78 &  60 & --- & --- & --- &  40 &  67 &  54 &  44 \\ 
FeI  &5956.706 & 0.86 & -4.34 &  63 & --- & --- & --- & --- &  48 & --- &
--- & --- & --- & --- &  40 & --- & --- \\ 
FeI  & 5969.578 & 4.28 &
-2.56 &   6 & --- & --- & --- & --- &   8 & --- & --- & --- & --- &
--- & --- & --- & --- \\ 
FeI  & 5983.688 & 4.55 & -0.62 &  74 & ---
& --- & --- & --- &  65 & --- & --- & --- & --- & --- &  60 & --- &--- \\ 
FeI  & 6003.022 & 3.88 & -0.90 &  92 & --- & --- & --- & ---
&  79 & --- & --- & --- & --- & --- &  76 & --- & --- \\ 
FeI  &6008.566 & 3.88 & -0.84 &  95 & --- & --- & --- & --- &  85 & --- &
--- & --- & --- & --- &  79 & --- & --- \\ 
FeI  & 6056.013 & 4.73 &
-0.39 &  80 &  63 & --- &  55 &  61 & --- &  60 & --- & --- & --- &
--- & --- & --- & --- \\ 
FeI  & 6078.499 & 4.79 & -0.20 &  89 &  71
&  75 &  56 &  67 & --- &  62 &  38 &  37 &  58 &  36 & --- &  64 & 55 \\ 
FeI  & 6079.016 & 4.65 & -0.88 &  54 &  40 &  45 &  28 &  38 &
--- &  36 &  17 &  16 &  31 &  16 & --- &  35 &  28 \\ 
FeI  &6082.708 & 2.22 & -3.48 &  40 &  27 &  33 &  29 &  32 & --- &  42 &  
8 &  10 &  16 &   7 & --- &  17 &  23 \\ 
FeI  & 6093.649 & 4.61 &
-1.35 &  32 &  26 &  26 &  18 &  20 & --- &  22 &   8 & --- &  12 &  
5 & --- &  18 &  11 \\ 
FeI  & 6096.671 & 3.98 & -1.80 &  39 &  27 & 
34 &  24 &  28 & --- &  30 &  11 &   9 &  18 &  10 & --- &  20 &  15\\ 
FeI  & 6098.250 & 4.56 & -1.76 &  18 &  14 &  13 &   8 &  11 &
--- &  13 & --- & --- & --- & --- & --- & --- & --- \\ 
FeI  &6151.623 & 2.18 & -3.27 &  52 &  38 & --- &  43 & --- &  48 &  55 & 
11 &  17 &  25 &  14 &  39 &  25 &  30 \\ 
FeI  & 6159.382 & 4.61 &
-1.89 &  13 &  10 & --- &   9 & --- &  16 &  10 & --- & --- & --- &
--- &  12 & --- & --- \\ 
FeI  & 6185.704 & 5.65 & -0.75 &  18 &  15
& --- &   9 & --- &  15 &   9 & --- & --- & --- & --- &  14 & --- &--- \\ 
FeI  & 6187.995 & 3.94 & -1.60 &  51 &  37 & --- &  33 & ---
&  45 &  41 & --- & --- & --- & --- &  37 & --- & --- \\ 
FeI  &6191.571 & 2.43 & -1.56 & 137 & 118 & --- & 116 & --- & 126 & 133 &
--- & --- & --- & --- & 118 & --- & --- \\ 
FeI  & 6696.322 & 4.83 &
-1.48 &  19 &  24 &  14 &  15 &  13 &  18 &  16 & --- & --- & --- &
--- &  11 & --- & --- \\ 
FeI  & 6699.136 & 4.59 & -2.10 &   9 &   8
&   7 &   5 &   7 &   9 &  11 & --- & --- & --- & --- &   4 & --- &--- \\ 
FeI  & 6703.576 & 2.76 & -3.01 &  39 &  26 &  33 &  28 &  34
&  36 &  44 & --- & --- & --- & --- &  27 & --- & --- \\ 
FeI  &6704.500 & 4.22 & -2.56 &   7 &   8 &   7 &   3 &   5 &   9 & --- &
--- & --- & --- & --- &   5 & --- & --- \\ 
FeI  & 6710.323 & 1.48 &
-4.79 &  17 &  10 &  15 &  14 &  16 &  13 &  28 & --- & --- & --- &
--- &  11 & --- & --- \\ 
FeI  & 6713.745 & 4.79 & -1.401 &  23 &  17
&  20 &   9 &  15 &  23 &  22 & --- & --- & --- & --- &  17 & --- &
--- \\ 
FeI  & 6725.364 & 4.10 & -2.17 &  19 &  11 &  16 &   9 &  12
&  16 &  16 & --- & --- & --- & --- &  10 & --- & --- \\ 
FeI  &6726.673 & 4.61 & -1.02 &  50 &  35 &  43 &  30 &  34 &  44 &  39 &
--- & --- & --- & --- &  37 & --- & --- \\ 
FeI  & 6730.307 & 4.91 &
-2.07 &   5 & --- &   6 & --- &   3 &   7 & --- & --- & --- & --- &
--- & --- & --- & --- \\ 
FeI  & 6732.068 & 4.58 & -2.13 &   9 &   7
&   7 & --- &   4 & --- &   8 & --- & --- & --- & --- &   4 & --- &
--- \\ 
FeI  & 6733.153 & 4.64 & -1.42 &  29 &  22 &  22 &  15 &  17
&  28 &  19 & --- & --- & --- & --- &  18 & --- & --- \\ 
FeI  &
6745.113 & 4.58 & -2.06 &  10 &  12 &   5 &   7 &   5 &  10 &   7 &
--- & --- & --- & --- &   8 & --- & --- \\ 
FeI  & 6745.984 & 4.07 &
-2.62 &   8 &   8 &   5 &   7 & --- &   9 &   5 & --- & --- & --- &
--- &   6 & --- & --- \\ 
FeI  & 6746.975 & 2.61 & -4.22 &   6 &   5
&   3 &   5 &   5 &   7 &   5 & --- & --- & --- & --- &   5 & --- &--- \\ 
FeI  & 6750.164 & 2.42 & -2.55 &  77 &  65 &  74 &  67 &  70
&  76 &  78 & --- & --- & --- & --- &  65 & --- & --- \\ 
FeII &5197.576 & 3.23 & -2.21 &  87 &  82 &  96 &  71 &  60 &  98 &  68 & 
74 & --- &  92 & --- & --- & --- & --- \\ 
FeII & 5234.630 & 3.22 &
-2.07 &  95 &  90 & 104 &  75 &  68 & 109 &  75 &  77 & --- &  99 & 
75 &  85 & 108 &  58 \\ 
FeII & 5264.812 & 3.33 & -2.82 &  57 &  51 &
 65 &  36 &  30 &  64 &  35 &  33 & --- &  55 &  31 &  51 &  62 &  29\\ 
FeII & 5991.378 & 3.15 & -3.57 &  33 & --- & --- & --- & --- & 
45 & --- & --- & --- & --- & --- &  28 & --- & --- \\ 
FeII & 6084.105& 3.20 & -3.81 &  22 &  22 &  31 &  14 &  10 & --- &  12 &  12 &   9
&  22 &  10 & --- &  31 &  10 \\ 
FeII & 6149.249 & 3.89 & -2.72 & 39 &  38 & --- &  23 & --- &  50 &  21 &  26 &  18 &  42 &  23 &  33 &
 50 &  14 \\  
CuI  &5782.136 & 1.64 &    --- &  82 &  48 &  71 &  64 &  66 &  74 &  82 &
--- & --- &  36 & --- &  60 &  31 &  37 \\ 
CuI  & 5218.209 & 3.82 &    --- &  60 &  44 &  57 &  40
&  45 &  61 &  47 &  20 &  19 &  34 &  19 &  43 &  34 &  25 \\
BaII & 5853.688 & 0.60 &
-0.76 &  68 &  67 &  79 &  64 &  50 &  79 &  66 & --- & --- &  71 & 
43 &  62 &  81 &  37 \\ 
BaII & 6141.727 & 0.70 &  0.30 & 125 & 119 &
--- & 108 & --- & 127 & 109 &  95 &  86 & 120 &  88 & 113 & 130 &  86\\
\end{tabular}
\end{table*}

\begin{table*}
\centering
  \caption{Hyperfine structure for CuI lines.}
  \begin{tabular}{@{}lccc@{}} 
Ion&$\lambda$($\rm \AA$)&$\chi_{ex}$& log{\it gf}\\[10pt]
CuI&5218.059&3.82&-1.33\\
CuI&5218.061&3.82&-0.85\\
CuI&5218.063&3.82&-0.98\\
CuI&5218.065&3.82&-0.26\\
CuI&5218.069&3.82&-0.48\\
CuI&5218.071&3.82&-0.48\\
CuI&5218.074&3.82&-0.14\\
CuI&5782.032&1.64&-3.53\\
CuI&5782.042&1.64&-3.84\\
CuI&5782.054&1.64&-3.14\\
CuI&5782.064&1.64&-3.19\\
CuI&5782.073&1.64&-3.49\\
CuI&5782.084&1.64&-2.79\\
CuI&5782.086&1.64&-3.14\\
CuI&5782.098&1.64&-3.14\\
CuI&5782.113&1.64&-2.79\\
CuI&5782.124&1.64&-2.79\\
CuI&5782.153&1.64&-2.69\\
CuI&5782.173&1.64&-2.34\\
\end{tabular}
\end{table*}

\begin{table*}
\centering
  \caption{Atmospheric parameters and elemental abundances for UMa
Group stars. HR2047, HR6094 and HR6748 were observed in both the
$\lambda$5810 and $\lambda$6145 regions: for these stars, the
[Ba/Fe] ratios refer to [FeII/H], for the other stars, no Fe II lines
were available and the [Ba/Fe] ratios refer to [FeI/H]. For all other
elements, the [element/Fe] ratios refer to [FeI/H]. The number of lines
available for each species is given in parentheses.}
\begin{tabular}{@{}lccccccc@{}} 
star&T$_{\rm eff}$ (K)&log g&$\xi$ (km.$\rm s^{-1}$)&\rm [FeI/H]&
\rm [FeII/H]
&\rm [Cu/Fe]&\rm [Ba/Fe]\\[10pt]

HR531B &5833&4.53&1.1 &-0.02(5) &--- &-0.12(1) &+0.39(1)\\
HR1321 &5557&4.51&0.9 &-0.27(6) &--- &-0.23(1) &+0.48(1)\\ 
HR1322 &5950&4.49&1.2 &+0.17(7) &--- &-0.12(1) &+0.32(1)\\ 
HR2047 &5929&4.49&1.1 &-0.02(10)&+0.10(2) &-0.14(1) &+0.29(2)\\
HD41593&5277&4.49&0.7 &+0.06(6) &--- &-0.22(1) &+0.39(1)\\
HR6094 &5895&4.52&1.5 &+0.03(13)&+0.07(2) &-0.11(1) &+0.30(2)\\
HR6748 &5895&4.49&1.0 &-0.06(9) &-0.03(2) &-0.17(1) &+0.44 (2)\\
\end{tabular}
\end{table*}

\begin{table*}
  \caption{Atmospheric parameters and elemental abundances for the
  normal disc stars.}
\begin{tabular}{@{}lccccccc@{}} 
star&T$_{\rm eff}$ (K)&log g&$\xi$ (km.$\rm s^{-1}$)
&\rm [Fe/H]&
\rm [Cu/Fe]&\rm [Ba/Fe]\\[10pt]

HR77 &5970&4.48&0.88 &-0.07     &-0.09(1) &+0.18(1)\\
HR98 &5860&4.05&1.50 &-0.11     &+0.07(1) &+0.03(1)\\ 
HR173 &5270&3.75&1.15&-0.70     &-0.06(1) &-0.02(1)\\ 
HR509 &5320&4.30&0.70&-0.50(2)  &-0.08(1) &-0.03(2)\\
HR695&5830&3.87&1.17 &+0.03     &-0.01(1) &+0.06(1)\\
HR914 &5020&3.66&0.73&-0.57(2)  &-0.13(1) &+0.01(2)\\
HR2883&5990&4.18&1.65&-0.75(2)  &-0.02(1) &-0.07 (2)\\
HR3018&5820&4.42&1.21&-0.78(2)  &-0.07(1) &-0.15 (2)\\
HR7875&5991&4.09&1.78&-0.44(2)  & 0.00(1) &-0.03 (2)\\
HR8181&6139&4.34&1.57&-0.67(2)  & 0.00(1) &-0.02 (2)\\
HR8501&5753&4.27&1.10&-0.25(2)  &-0.04(1) &+0.01 (2)\\
HR8697&6288&3.97&2.17&-0.25(2)  &-0.03(1) &+0.07 (2)\\
HR9088&5551&4.45&0.85&-0.73(2)  &-0.13(1) &-0.18 (2)\\
\end{tabular}
\vspace{1cm}
\end{table*}

\begin{table*}
  \caption{Dependence of Cu and Ba abundances on input parameters.}
\begin{tabular}{@{}lccccc@{}} 
&$\rm \Delta T_{eff}$&$\rm \Delta log g$&$\rm \Delta$ [Fe/H]&
$\rm \Delta \xi$ & $\rm \Delta W_\lambda$ \\
& +100 K & $-$0.10 cm.$\rm s^{-2}$ & +0.10 dex & +0.3 km.$\rm s^{-1}$ &
+3 m\AA \\[10pt]

\rm [Cu/Fe] & +0.06 & +0.01  & +0.01 & $-$0.01 & +0.06\\
\rm [Ba/Fe] & +0.03 & $-$0.02& +0.04 &$-$0.09& +0.04 \\ 
\end{tabular}
\end{table*}

\section{Conclusions and Discussion}

One of the objectives of this analysis was to obtain Ba abundances of
the Ursa Major Group and therefore verify the status
of the member HR6094 as a Ba dwarf star. The anomalous Ba dwarf status of 
HR6094 was
previously proposed by Porto de Mello \& da Silva (1997) as being
unique for a young, solar metallicity, though we note that Jeffries \&
Smalley (1996) report a solar metallicity,
rapidly rotating K-dwarf star, member of a binary system, which also
shows Ba excess.
Figure 1 shows the observed spectra of the UMaG member, HR6748, and of
a normal star, HR77. The normal star's atmospheric parameters are very
similar
to those of HR6748 as seen in Tables 4 and 5. It is clearly seen that the
$\lambda$5853 BaII line is stronger in the UMaG star spectrum. 
Since the [Ba/Fe] abundance ratio have shown to change +0.03 as 
T$_{\rm eff}$ changes
+100 K, one would expect the $\lambda$5853 BaII line to be stronger in
HR77 spectrum which is slightly hotter than HR6748, if the two stars had
a similar [Ba/Fe] ratio.

\begin{figure}
\protect\centerline{
\epsfxsize=8cm
\epsfbox{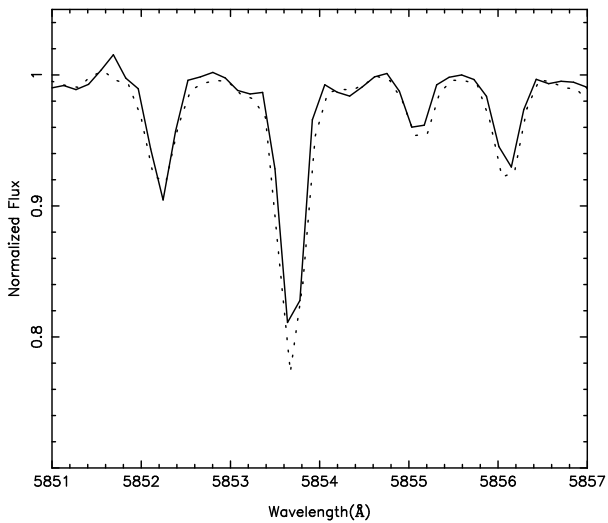}}
\caption{Observed spectra of the normal disc star, HR77
(full line) and
the Ursa Major Group member, HR6748 (dotted line). The $\lambda$5853 BaII 
feature is stronger for HR6748 than for HR77. The other three features
are $\lambda$5852, $\lambda$5855 and $\lambda$5856 Fe I lines.}
\label{fig:allcont}
\end{figure}

The observed [Ba/Fe] for the UMaG stars stand out
in their comparison to solar neighborhood normal stars with the same
metallicity. Figure 2 shows [Ba/Fe] vs. [Fe/H] for the UMaG stars,
plotted with normal disc stars from E93 and the stars listed in Table
5. We may assume good homogeneity between our abundance data and
E93's, as judged by the similar methods of analysis and the very
similar metallicities found for the common stars (section 2):
particularly, for HR2047 the atmospheric parameters, Fe and Ba
abundances show excellent agreement with our values. One can see that 
the UMaG stars (represented by filled circles in the figure) form a separated
group resembling the locus of Ba dwarf stars represented by crosses
in the figure. 

\begin{figure}
\protect\centerline{
\epsfxsize=8cm
\epsfysize=7cm
\epsfbox{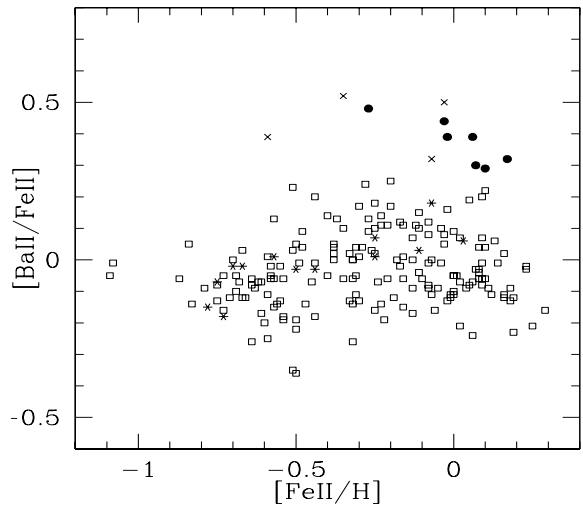}}
\caption{ [BaII/FeII] ratio versus [FeII/H] for the Ursa Major Group 
stars (filled circles), disc stars from Table 5 (stars), disc
stars (open squares) and Ba dwarfs (crosses) from Edvardsson 
{\it et al.} (1993).}
\end{figure}

\begin{figure}
\protect\centerline{
\epsfxsize=5cm
\epsfysize=5cm
\epsfbox{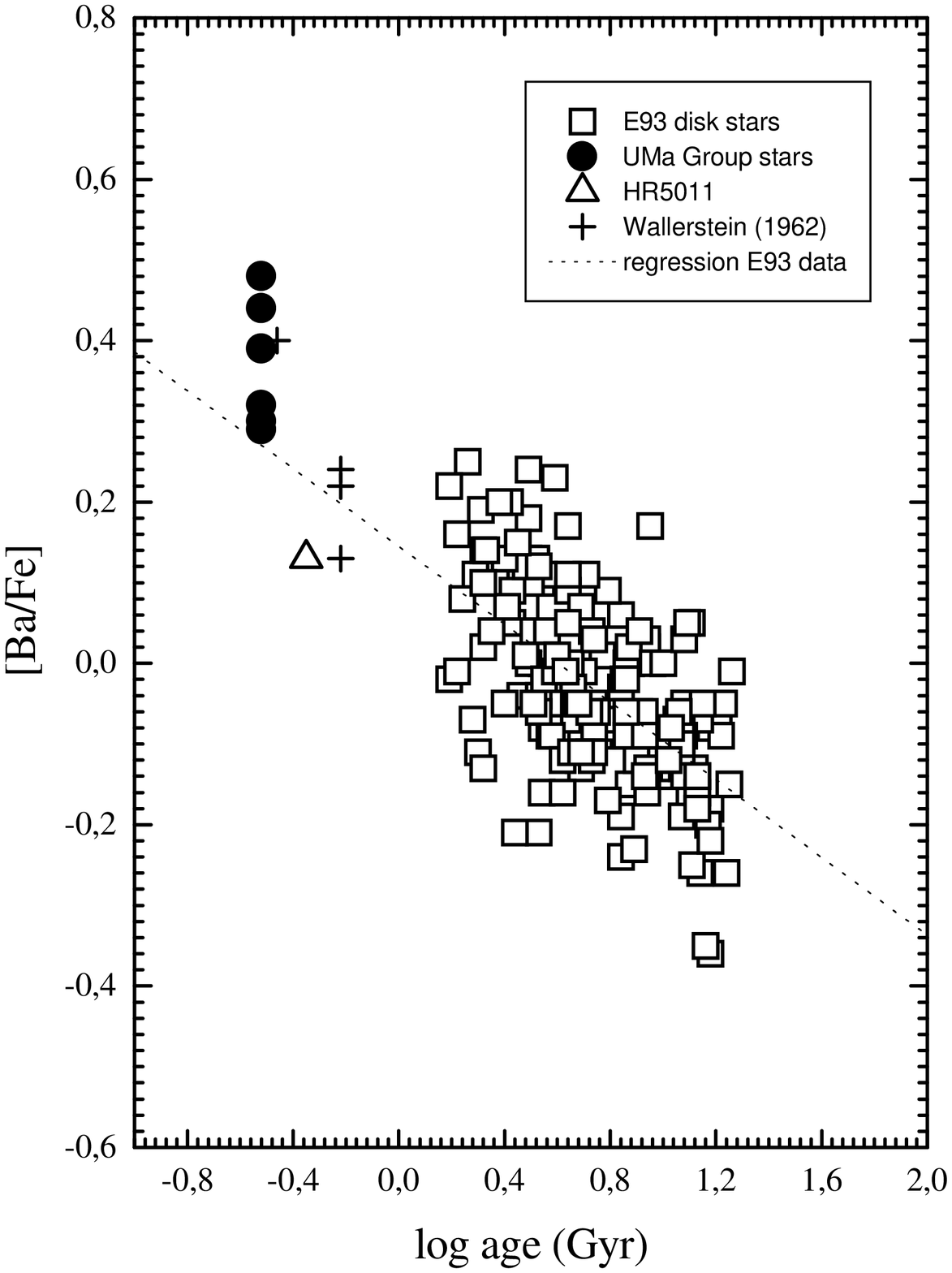}}
\caption{ [BaII/FeII] ratio versus log age (Gyr) for the UMaG 
stars (filled circles); another UMaG member HD115043 given in
Wallerstein (1962) (upper cross); disc stars from 
Edvardsson {\it et al.} (1993) (open squares); a young star, HR5011 
(triangle) and
Hyades stars from Wallerstein (1962) (the three lower crosses). The regression 
line corresponds to Edvardsson et al. (1993)'s points only.}
\end{figure}

There are a few other
UMaG stars with abundance determinations in the literature. For
those stars with derived Ba abundances, there is a good agreement in
the existence of an overabundance of Ba relative to Fe.
The study by Wallerstein (1962) found for HD115043 (a probable
member of UMaG according to Soderblom \& Mayor 1993) 
[Ba/Fe]=+0.40 based on photographic plates.
Oinas (1974) performed a model atmosphere analysis of $\xi$ Boo (a
possible UMaG member) obtaining [Ba/Fe]=+0.2 based also on a
photographic plate. These studies claim uncertainties of
$\sim$0.2 dex in their analyses. More
recently, E93 analyzed HR2047, another UMaG member also
analyzed here,
and found [Ba/Fe]=+0.25 and [Y/Fe]=+0.31, while Porto de Mello
\& da Silva (1997) found [Ba/Fe]=+0.37 and [Y/Fe]=+0.22 for HR6094,
which they thought was a Ba dwarf. 
E93 argue that their Ba abundance determination
might be considered normal for young stars in the light of an inverse
[Ba/Fe] correlation with age.


In Figure
3 we have plotted the [Ba/Fe] versus age relation from E93 and added
points for: the present results for the UMaG stars; HD115043 (UMaG
member) and three Hyades stars from Wallerstein (1962); and a young
normal disc star, HR5011, from Porto de Mello (1998). The stars from
E93 had their ages determined from theoretical isochrone fitting to
observational HR diagrams and are all older than about 1.6 Gyr, a
measure taken to increase age determination reliability. The UMaG and
the Hyades have well known ages of 0.3 Gyr and 0.6 Gyr, respectively.
HR5011 has chromospheric activity level (log R'$_{\rm HK}$ = $-$4.43, from
Soderblom 1985)
compatible with both the UMaG stars ($<$ log R'$_{\rm HK}$ $>$ = $-$4.4, from
Soderblom \& Mayor 1993), and the Hyades stars ($<$ log R'$_{\rm HK}$
$>$ = $-$4.3, from Soderblom \& Clements 1987). Since the ability of the
chromospheric activity level to distinguish ages much below 1 Gyr is
considerably reduced (Soderblom \& Clements 1987), we have taken the
age of HR5011 to be intermediate between the UMaG and the Hyades. We can 
basically draw two 
hypothesis from Figure 3:
a) the UMaG stars are in the upper tail of the Ba-age relation because
they are young as concluded by E93 for HR2047; b) the Ba excess is anomalous 
and stands out as compared to
normal stars suggesting a primordial origin for the UMaG abundances.
In any of these cases, the fact that all analyzed stars in the group
show overabundances of Ba is striking. The enrichment of s-process elements from an AGB star through
the Ba star phenomenon could be a reasonable explanation {\it if not}
all the seven stars had shown similar overabundances.
There is an insufficient number of stars younger than 1.5 Gyr in Figure 3
to permit any further conclusion. Certainly, such
lack of stars postpones any conclusion about what is going on with the Ba
abundances for the youngest stars in the solar vicinity.
However the uncertainties in the abundance ratios of the sample stars are 
low enough to establish the accuracy of the observed Ba abundances.

The present Ba abundances lead us to believe that HR6094 does not seem 
to be a Ba dwarf but indeed a member of
a kinematical group in which {\bf all} analyzed member stars are enriched
in Ba. The present results put HR6094 rather as a member of an 
anomalously
Ba-enriched group of stars, no longer supporting its Ba dwarf status.
The Ba star phenomenon cannot account for the anomalies
found in UMaG stars, since it seems unlikely that
an AGB star has enriched all sample stars at the same time
in a distance scale of tens of parsecs. Porto de Mello \& da Silva
(1997) have found abundance anomalies for other elements in HR6094.
Figure 4 shows the abundance pattern for HR6094 in which the
s-process elements (Y, Zr, Ba, Ce, Pr and Nd) all have abundances
above solar. The most striking overabundance is clearly for Ba,
which also has the largest error bar ($\sim \pm$0.1). The overabundance 
of s-process elements in the diagram is real within the uncertainties.

Other suggested anomalies are Na, C and Cu deficiencies
with respect to normal solar-metallicity stars. The average [Na/Fe]
abundance ratio found for three UMaG stars with available NaI
lines is $-$0.15 dex.
Porto de Mello \& da Silva (1997) and Tomkin, Woolf \& Lambert (1995)
respectively, have found a 0.2 dex C-deficiency for both
HR6094 and HR2047. As argued by Porto de Mello \& da Silva,
a C-deficiency may accompany the operation of the hot-bottom
burning in AGB stars. Regarding the homogeneity of the abundance
pattern found for the analyzed member stars, it is
likely that the Group as a whole is C-deficient.

\begin{figure}
\protect\centerline{
\epsfxsize=5cm
\epsfysize=5cm
\epsfbox{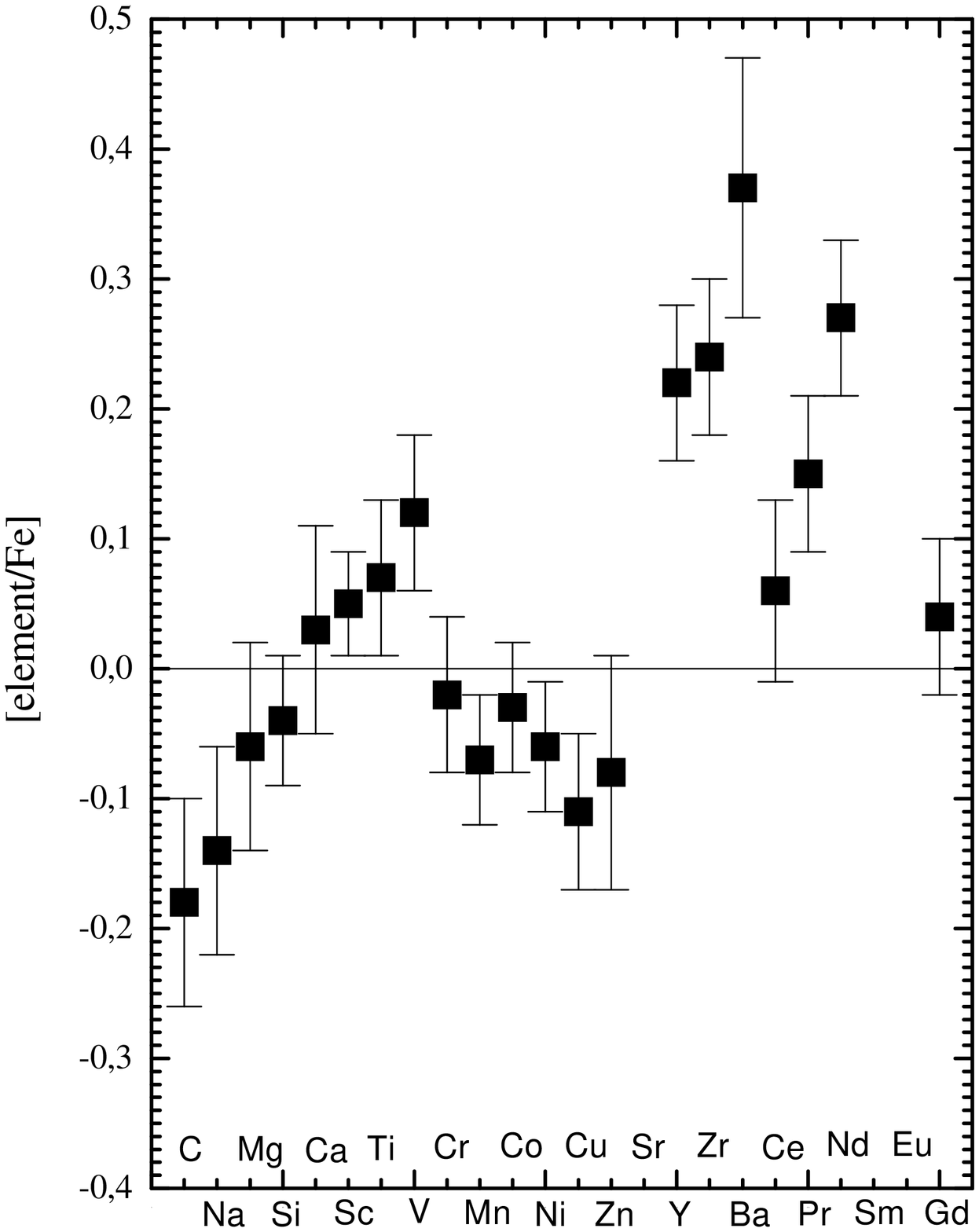}}
\caption{Abundance pattern for the Ursa Major Group member HR6094.}
\end{figure}

The results
found for the Cu abundances deserve a special attention as is shown in
Figure 5 for the [Cu/Fe] abundance ratios. We have plotted [Cu/Fe] vs.
[Fe/H] for the UMaG stars and disc stars, as found in the present
work, and stars from Sneden {\it et al.} (1991) and Porto de Mello (1998),
that lie in the same metallicity range. Porto de Mello's (1998)
analysis follows strictly the same methods employed in the present
work, while Sneden {\it et al's} (1991) data are somewhat heterogeneous
particularly in what pertains to the atmospheric parameters. Keeping
in mind these limitations, nevertheless the UMaG stars appear as an
isolated group as compared to normal stars with the same [Fe/H].
The [Cu/Fe] ratio increases as [Fe/H] increases for the normal stars.
Unfortunately, there are no Cu abundances determined for other stars
with [Fe/H]$>$ 0, although the turnup in the [Cu/Fe] versus [Fe/H]
diagram is clear at [Fe/H]$\sim$0 (marked by the vertical dotted line). 
Since SNeIa starts to produce
iron-peak elements at [Fe/H]$\sim-$1.2 (Matteucci {\it et al.} 1993),
the [Cu/Fe] ratio should be constant with [Fe/H] after 0.0. It
rather increases for higher [Fe/H] which might suggest that there
is another nucleosynthetic process producing Cu.

The [Cu/Fe] deficiency seems
to follow the [Ba/Fe] overabundance, i.e., [Cu/Fe] decreases as
the [Ba/Fe] ratio increases, suggesting that there may be a
connection between the Cu destruction and the production of Ba
(and other heavy elements produced by the main component of
the s-process). Another piece of evidence in favor of a
nucleosynthetic connection between the s-process and Cu has been
recently found by Pereira \& Porto de Mello (1997) and Pereira {\it et al.}
(1998), who found remarkable Cu depletions for two Ba stars which are
also symbiotic systems: both appear appreciably enriched in the
s-process elements, resembling classical low metallicity Ba giants,
and present remarkable Cu deficiencies with respect to giant halo
stars of the same metallicity. Figure 6 shows a plot of [Cu/Fe] versus
[Ba/Fe] for the UMaG stars, disc stars and these two Ba stars. The
fact that Cu is deficient in these Ba stars, where the s-process
elements are enhanced, leads us to believe that Cu is depleted in the
process of synthesizing neutron capture elements. Another evidence for
this anti-correlation between Cu and Ba comes from stars more
metal-rich than the Sun. For higher metallicities the [Cu/Fe] ratio is
supersolar (Figure 5). If one considers that Castro {\it et al.} (1997) and
E93 found that the [Ba/Fe] ratio is underabundant for disc stars with
[Fe/H] $>$ 0, a possible conclusion would be that as Ba becomes
underabundant, the abundance of Cu goes up, indicating that Cu may be
acting as seed for the neutron capture process, and its depletion
would be lessened in the metal-rich stars, which are underabundant in
Ba. Unfortunately these last two studies do not provide Cu abundances.
Clearly, it is necessary to obtain further Cu abundances for stars
with [Fe/H] $>$ +0.3: a comprehensive study of the connection of Ba and
Cu abundances in normal disc stars with -1.0 $<$ [Fe/H] $<$ +0.3 and Ba
stars is presently underway.

\begin{figure}
\protect\centerline{
\epsfxsize=8cm
\epsfysize=7cm
\epsfbox{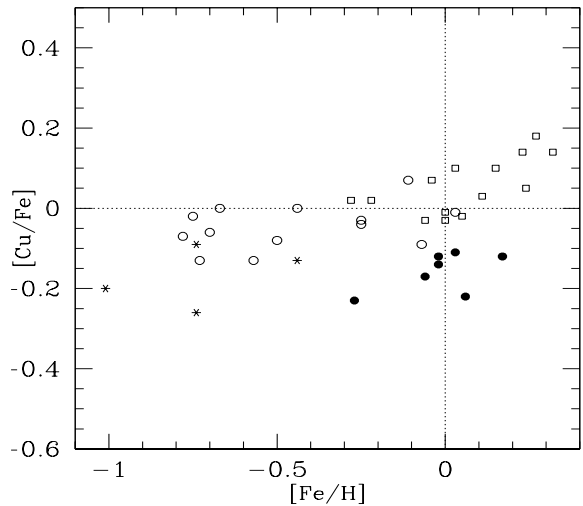}}
\caption{[Cu/Fe] ratio versus [Fe/H] for the Ursa Major Group stars
(filled circles) and disc stars from: Table 5 (open circles),
Porto de Mello (1998) (open squares) and
Sneden {\it et al.} (1991) (only the stars in the same metallicity range) 
(stars).}
\end{figure}

\begin{figure}
\protect\centerline{
\epsfxsize=8cm
\epsfysize=7cm
\epsfbox{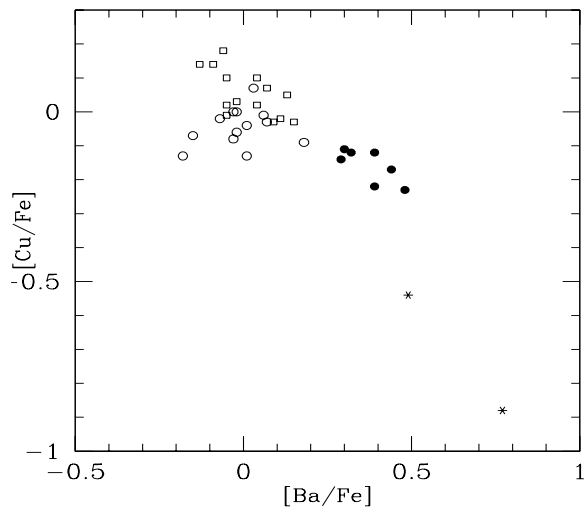}}
\caption{[Cu/Fe] ratio versus [Ba/Fe] for the Ursa Major Group stars
(filled circles); disc stars from: Table 5(open circles) and
Porto de Mello (1998) (open squares); the symbiotic star He 2-467 from
Pereira {\it et al.} (1998) and the symbiotic \& Ba star
BD-21$^\circ$ 3873 from Pereira \& Porto de Mello (1997) (stars).}
\end{figure}

The hypothesis of a primordial origin for the abundance anomalies
of the UMaG may be considered, i.e., perhaps the original cloud from which
the UMaG stars were formed was enriched in s-process
elements and depleted in Ca, Na and Cu. The most likely explanation
for the anomalous abundance pattern of the UMaG stars is thus a
common origin in a chemically-peculiar giant molecular
cloud. The present results therefore suggest that inhomogeneities
are present in the interstellar medium, and that localized
chemical evolution events do affect the abundance pattern
observed in disc stars. Boesgaard, Budge \& Burck (1988) have argued
that the mixing of enriched material from which the cluster
stars were made was not uniform about 0.3 -- 0.7 Gyr ago
for the young star clusters in the solar vicinity.
One could hypothesize that the ejecta of a few AGB stars,
early in the lifetime of the group, are responsible for the observed
altered Cu and s-process abundances. This event need not to have happened
necessarily during the condensation of the group, but possibly early
enough that it still retained high spatial cohesion. It remains highly
speculative that appreciable enrichment through matter accretion could
remain efficient over distance scales of tenths of parsecs, but it is
not inconceivable that AGB stars with 4--5 M$_\odot$ may have evolved
fast enough to provoke accretion of nucleosynthetically altered matter
upon their lower mass fellow group members before the group lost its
initial compactness. Such a scenario will be tested when accurate and
detailed abundance data of other young stars of the solar vinicity
become available, to be compared with the abundance pattern found here
for the UMaG. One may then decide if the UMaG is either indeed
anomalous or merely reflects the chemical evolution of the galactic
disc for very young star systems. Possibly, the remnants of those AGB
stars are now observed as the two white dwarf stars that
follow the member stars HR6094 and Sirius.

We can thus
summarize the chemical pattern of the UMaG stars as follows: they are
C-deficient by 0.2 dex, Na-deficient by 0.15 dex, Cu-deficient
by 0.2 dex and enriched in the main s-process elements by
0.3 -- 0.5 dex. We have
already launched an observational effort to obtain echelle spectra for
all the probable UMaG members in order to investigate the elements
that have shown to be anomalous in the Group. Obviously, further
data on the abundance pattern of young clusters and kinematical
groups are of great interest.

\section{Acknowledgments}

S.C acknowledges CNPq/Brazil research grant 300003/97-8. We thank
FUJB, FAPERJ/Brazil and PRONEX/Finep n. 41.96.0908.00 for 
financial help. We are deeply
indebted to all the staff of the CNPq/Observat\'orio do Pico dos Dias
for renewed helpfulness during the observations collected for this
work. We are grateful to C. Sneden for discussions concerning the
nucleosynthesis of Cu. The authors thank the referee R. D. Jeffries
for useful comments.

{}

\begin{thebibliography}{}
\bibitem{} Boesgaard A. M., Budge K. G., Burck E. E., 1988, ApJ, 325, 749
\bibitem{} Castro S., Rich R. M. Grenon M., Barbuy B., McCarthy J. K.,
1997, AJ, 114, 376
\bibitem{b1} Charbonnel C., Meynet G., Maeder A., Schaller D., 1993, A\&A,
101, 415
\bibitem{b2} Edvardsson B., Andersen J., Gustafsson B., Lambert
D. L., Nissen P. E., Tomkin J., 1993, A\&A, 275, 101
\bibitem{b3} ESA, 1997, The Hipparcos Catalogue, ESA SP-1200
\bibitem{} Jeffries R. D., Smalley B., 1996, A\&A, 315, L19
\bibitem{b4} Habets G. M. J., Heintze J. R. W., 1981, A\&AS, 46, 193
\bibitem{b5} Kurucz R. L., Furenlid I., Brault J., Testerman L.,
1984, The Solar Flux Atlas from 296 nm to 1300 nm, National Solar
Observatory
\bibitem{} Matteucci F., Raiteri C. M., Busson M., Gallino R., Gratton
R., 1993, A\&A, 272, 421
\bibitem{b6} Meylan T., Furenlid I., Wiggs M. S., 1993, ApJS, 85,
163
\bibitem{b7} Oinas, V., 1974, ApJS, 27, 405
\bibitem{b8} Pereira C. B., Porto de Mello G. F., 1997, AJ, 114, 2128
\bibitem{b9} Pereira C. B., Smith V. V, Cunha K. M. L., 1998, submitted
\bibitem{b10} Porto de Mello G. F., 1998, submitted
\bibitem{b11} Porto de Mello G. F., da Silva L., 1997, ApJL, 476, L89
\bibitem{b12} Saxner M., Hammarb\"ack G., 1985, A\&A, 151, 372
\bibitem{} Sneden C., Gratton R. G., Crocker D. A., 1991, A\&A 246, 354
\bibitem{} Soderblom D. R., 1985, AJ, 90, 2103
\bibitem{} Soderblom D. R., Clements S. D., 1987, AJ, 93, 920
\bibitem{b13} Soderblom D., Mayor M., 1993, AJ, 105, 226
\bibitem{b14} Steffen M., 1985, A\&AS, 59, 403
\bibitem{b15} Tomkin J., Woolf V. M., Lambert D. L., 1995, AJ, 109,
2204
\bibitem{} Vanture A. D., 1992, AJ, 104, 1997
\bibitem{b16} Wallerstein G., 1962, ApJS, 61, 407
\end{thebibliography}
\end{document}